# P–vortices and Drama of Gribov Copies.

V.G. Bornyakov[a,b], D.A. Komarov[b] and M.I. Polikarpov[b]

[a] *Institute for High Energy Physics, Protvino 142284, Russia*
[b] *Institute of Theoretical and Experimental Physics,*
*B.Cheremushkinskaya 25, Moscow, 117259, Russia*

**Abstract**

We present results of the careful study of the Gribov copies problem in $SU(2)$ lattice gauge theory for the direct maximal center projection widely used in confinement studies. Applying simulated annealing algorithm we demonstrate that this problem is more severe than it was thought before. The projected (gauge noninvariant) string tension is not in the agreement with the physical string tension. We do not find any indications that P–vortices reproduce the full $SU(2)$ string tension neither in the infinite volume limit nor in the continuum limit.

# 1 Introduction

The idea that the center vortices are the objects responsible for confinement in the non-abelian gauge theories is rather old [1, 2]. Recently it has been argued that the center projection might provide a powerful tool to investigate this idea [3]. It is suggested that projection dependent P-vortices defined on the lattice plaquettes are able to locate thick gauge invariant center vortices and thus provide the essential evidence for the center vortex picture of confinement. So far 3 different center gauges have been used in practical computations: the indirect maximal center (IMC) gauge [3], the direct maximal center (DMC) gauge [4] and the Laplacian center gauge [5] (see also [6] for a new proposal).

IMC and DMC gauges suffer from the Gribov copies problem [7] and this technical problem gave rise to several claims and counterclaims. At present we have a drama in five acts:

*i)* Initially there was the claim [3, 4] that the projected string tension (the string tension which is due to P-vortices) reproduces the full $SU(2)$ string tension. Thus P-vortices are responsible for the confinement of color in gluodynamics.

*ii)* The problem of Gribov copies has been raised in Ref. [8], where it has been demonstrated that there are gauge copies which produce P-vortices evidently with no center vortex finding ability since projected Wilson loops have no area law. At the same time these gauge copies correspond to higher maxima of the gauge fixing functional, than those used in [4].

*iii)* In our previous publications [9, 10] we confirmed the existence of two classes of gauge copies: with nonzero and zero projected string tension $\sigma_{Z(2)}$. We resolved the problem raised in [8] since we found the copies with the highest maxima of the gauge fixing functional which correspond to nonzero $\sigma_{Z(2)}$. At the same time the value of $\sigma_{Z(2)}$ was essentially lower than that obtained in [4] and in is disagreement with the physical string tension $\sigma_{SU(2)}$.

*iv)* In Refs. [11, 12] it has been argued that the disagreement between $\sigma_{Z(2)}$ and $\sigma_{SU(2)}$ is due to strong finite volume effects and it disappears on large enough lattices. In Refs. [9, 10, 11, 12] the usual relaxation plus overrelaxation (RO) algorithm was used to fix the DMC gauge.

*v)* Below we present results[1], obtained with the help of more powerful gauge fixing algorithm, simulated annealing (SA), and show that the problem of low value of $\sigma_{Z(2)}$ persists even on large lattices with physical extension up to $3fm$. We discuss here DMC gauge. Our results for IMC gauge will be presented elsewhere.

# 2 Gauge fixing procedure

## 2.1 Gribov copies

Direct maximal center gauge [4] in $SU(2)$ lattice gauge theory is defined by the maximization of the following functional:

$$F(U) = \frac{1}{4V}\sum_{n,\mu}\left(\frac{1}{2}\mathrm{Tr}U_{n,\mu}\right)^2 = \frac{1}{4V}\sum_{n,\mu}\frac{1}{4}\left(\mathrm{Tr}_{adj}U_{n,\mu}+1\right), \qquad (1)$$

with respect to local gauge transformations, $U_{n,\mu}$ is the lattice gauge field, $V$ is the lattice volume. Condition (1) fixes the gauge up to $Z(2)$ gauge transformation, and can be considered as the Landau gauge for adjoint representation. Any fixed configuration can be decomposed into $Z(2)$ and coset parts: $U_{n,\mu} = Z_{n,\mu}V_{n,\mu}$, where $Z_{n,\mu} = \mathrm{sign}\mathrm{Tr}U_{n,\mu}$. The plaquettes $Z_{n,\mu\nu}$ constructed from the links $Z_{n,\mu}$ have values $\pm 1$. The P-vortices (which form closed surfaces in 4D space) are made from the plaquettes, dual to plaquettes with $Z_{n,\mu\nu} = -1$.

[1] The preliminary results are presented in Ref. [10]

Some evidence has been collected, that P-vortices in DMC gauge can serve to locate gauge invariant center vortices. It has been reported [4] that the expectation value of the projected Wilson loops $W_{Z(2)}$, computed via linking number of the static quarks trajectories and P-vortices, have area law. The projected string tension $\sigma_{Z(2)}$ is very close to the string tension of the nonabelian theory $\sigma_{SU(2)}$. This fact has been called center dominance. Another important observation was that the density of P-vortices scales as a physical quantity [4, 13].

The main problem of DMC gauge fixing is that the functional $F(U)$ (1) has many local maxima. This is the analogue of the Gribov problem in continuum gauge theories [7] and we call configurations corresponding to these local maxima Gribov copies. For gauge conditions with Gribov copies the usual expression for gauge dependent quantities [14] does not provide unambiguous definition (see recent discussion of this subject in [15]). It was shown that for some gauge conditions the gauge dependent quantities depend strongly on the local maxima picked up [16, 17]. Thus to remove this ambiguity it is necessary to find the global maximum or, if this is impossible, to approach the global maximum as close as possible[2].

The usual algorithm for the gauge of the type (1) is the relaxation algorithm which performs maximization iteratively site by site. The relaxation becomes more effective if one uses the overrelaxation. It is also known that the SA algorithm is more effective than RO algorithm. It is very useful when ambiguities induced by Gribov copies become severe [17, 19]. For this reason we use this algorithm for DMC gauge.

We follow procedure proposed and checked in [17]: for given configuration we generate $N_{cop}$ gauge equivalent copies applying random gauge transformations to the initial configuration, after that we fix the gauge for each gauge copy thus producing $N_{cop}$ Gribov copies for each configuration. Then we compute a gauge dependent quantity $X$ on the configuration corresponding to the Gribov copy with the highest local maximum of (1), $F_{max}(N_{cop})$. Averaging over statistically independent gauge field configurations and varying $N_{cop}$ we obtain the function $< X(N_{cop}) >$ and extrapolate it to $N_{cop} \to \infty$ limit. This should provide a good estimation for $< X >$ computed on the global maximum unless the algorithm in use produces local maximum far from the global ones.

Note that there exists another proposal [20] for the gauge fixing procedure which is free of gauge copies problem. In some particular limit this procedure corresponds to the search of the global maximum. There is also a class of gauge conditions [21, 5, 6] which do not suffer from the gauge copies problem.

[2]For another proposal to solve this problem see e.g. [18].

## 2.2 Simulated annealing algorithm

The functional $F(U)$ can be considered [22] as the spin action:

$$S(s) = F(U^g) = \frac{1}{16V} \sum_{n,\mu} \mathrm{Tr}_{adj} \left( s_n U_{n,\mu} s_{n+\hat{\mu}} + 1 \right) \tag{2}$$

where the "spin variable" $s_n$, is the gauge transformation matrix in the adjoint representation: $s_{n,ab} = \frac{1}{2} tr(\sigma_a g_n \sigma_b g_n^\dagger)$ .

The lattice field $U_{n,\mu}$ plays the rôle of (almost) random local couplings. Maximization of the functional $F(U^g)$ is equivalent to decreasing the auxiliary temperature $T$ of the statistical system with the partition function,

$$Z = \sum_{\{s_n\}} \exp \left( \frac{1}{T} S(s) \right) \quad . \tag{3}$$

One starts with equilibrating this spin glass at high temperature. Subsequently the temperature, $T$, is decreased adiabatically. In the limit $T \to 0$ the system approaches the ground state, i.e. the maximal value of $S(s)$. Our SA implementation consists of the following three steps:

1. Thermalization at $T = 1.16$.

2. Gradual decreasing of $T$ down to $T = 0.02$.

3. Final maximization by means of the RO algorithm.

In steps 1 and 2 an updating of the spin configuration was done with Metropolis algorithm. The temperature was lowered after every sweep by the quantity $\delta T(T)$. $\delta T(T)$ has been tuned such that the spin action increased about linearly with the number of iteration sweeps. This has been realized by subdividing full range [0.02,1.16] into 38 intervals of width $\Delta T = 0.03$. The corresponding differences of the average action $\Delta \bar{S}(T) >= \bar{S}(T) - \bar{S}(T - \Delta T)$ have been computed on equilibrated configurations and were found to be very stable against statistical fluctuations among different Monte Carlo configurations. We used the same $\delta T(T)$ for all volumes at given value of gauge coupling $\beta$ and slightly modified it for different $\beta's$. The number of sweeps to be performed within each interval $(T - \Delta T, T]$ was chosen to be proportional to $\Delta \bar{S}(T)$ and, subsequently, the corresponding value of $\delta T(T)$ has been determined. To approach close enough to the equilibrium we have to perform about 2000 sweeps at the step 2 (for step 1 only 20 sweeps was enough). Finally, at step 3 it is necessary to make $O(100)$ sweeps of the RO algorithm to satisfy convergence criterion. In total this procedure was essentially more costly than the RO algorithm, also it gives the higher maximum of the functional (1).

# 3 Results

## 3.1 Details of simulation

Our computations have been performed on lattices $12^4, 16^4, 20^4$ and $24^4$ for $\beta = 2.4$ and $16^4, 20^4$ for $\beta = 2.5$. The number of configurations was 150 for $12^4$, 50 for $16^4$, 30 for $20^4$, 20 for $24^4$.

Using the described above gauge fixing procedure we calculate various observables as functions of the number of randomly generated gauge copies $N_{cop}$. To make computations feasible we reduce the maximal value of $N_{cop}$ on large lattices, see Table 1.

Table 1: The maximal values of $N_{cop}$ for various lattices.

| | $L = 12$ | $L = 16$ | $L = 20$ | $L = 24$ |
|---|---|---|---|---|
| $\beta = 2.4$ | 10 | 7 | 7 | 4 |
| $\beta = 2.5$ | – | 7 | 7 | 7 |

The projected Wilson loops $W_{Z(2)}(C)$ are defined as

$$W_{Z(2)}(C) =< \exp\{i\pi\mathcal{L}(\Sigma_P, C)\} >$$

where $\mathcal{L}(\Sigma_P, C)$ is the 4D linking number of the closed surface, $\Sigma_P$, formed by P-vortex and closed loop $C$. We used two ways to estimate projected string tension $\sigma_{Z(2)}$: from the Creutz ratios

$$\chi_{Z(2)}(I, I) = -\log\frac{W_{Z(2)}(I, I)W_{Z(2)}(I+1, I+1)}{W_{Z(2)}(I, I+1)W_{Z(2)}(I+1, I)}\,; \tag{4}$$

and fitting static potential

$$V_{Z(2)}(R) = -\lim_{T\to\infty}\log\frac{W_{Z(2)}(R, T+1)}{W_{Z(2)}(R, T)}$$

to the usual form

$$V_0 + c/R + \sigma_{Z(2)}a^2R\,. \tag{5}$$

The P-vortex density $\rho$ is another quantity of interest. It is defined as follows

$$\rho = \frac{1}{12\cdot V}\sum_{n;\mu>\nu}(1 - Z_{n,\mu\nu}).$$

### 3.2 Maximized functional and extrapolation to $N_{cop} \to \infty$

In DMC gauge the average value $< F_{max} >$ is an important indicator of the quality of the gauge fixing since gauge noninvariant observables we are measuring are strongly correlated with $< F_{max} >$. In Fig.1 we plot $< F_{max} >$ as a function of $N_{cop}$. For comparison we show the results for RO algorithm and the algorithm suggested in [8] : before applying RO algorithm the random gauge copy was fixed to Landau gauge. We call it LRO algorithm. One can see that SA algorithm indeed produces highest maxima. We checked that this is true for other lattices as well.

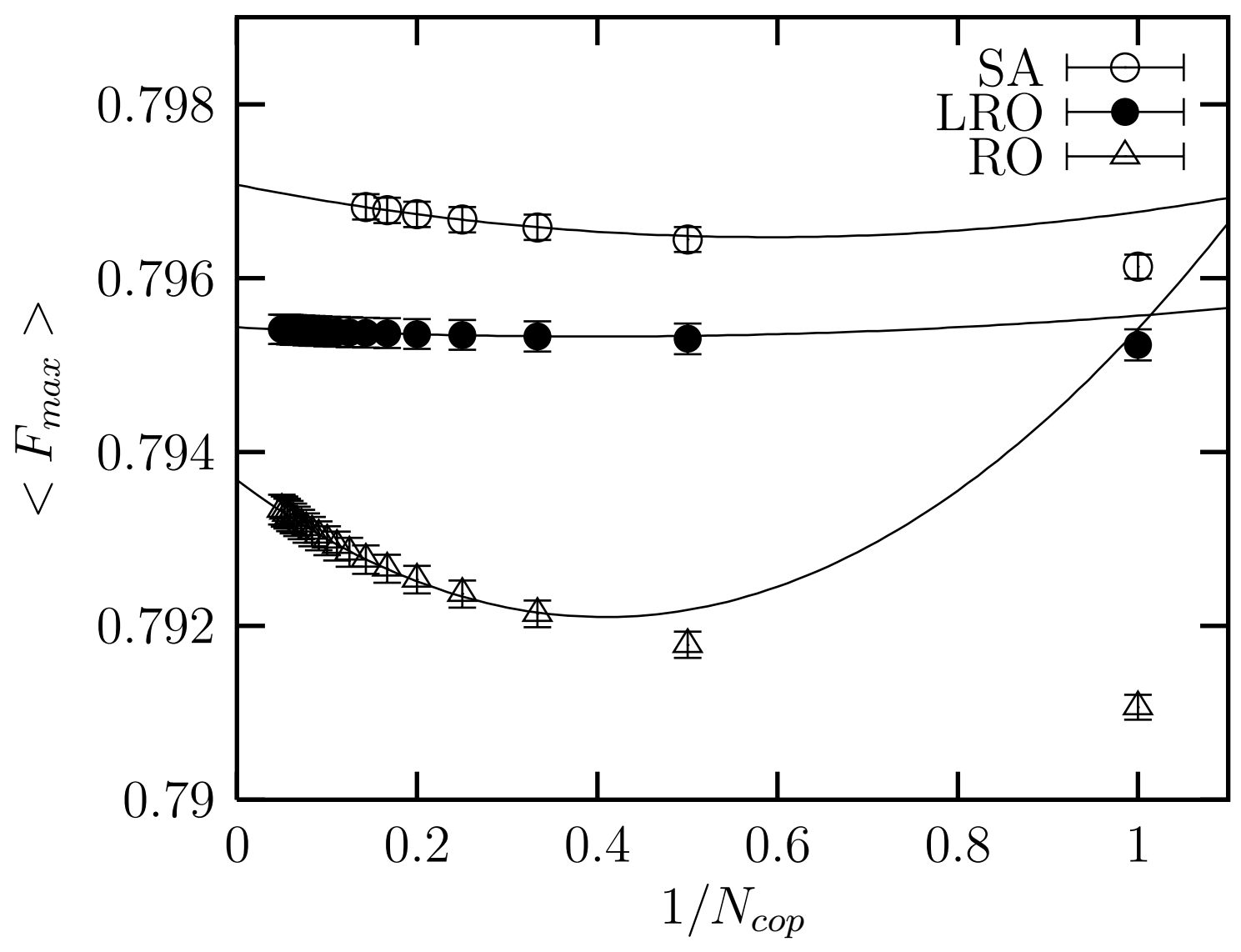

Figure 1: The average value of the highest local maxima $< F_{max} >$ for 3 algorithms as a function of $N_{cop}$; $L = 16, \beta = 2.5$. The solid lines are the fits using the asymptotic expression (A.1).

One can draw two conclusions from Fig.1. First, it is clear that the local maxima generated with LRO procedure of [8] can be safely ignored. Second, RO algorithm does not permit to reach the global maxima and results obtained with this algorithm must be revised with the help of SA algorithm. As we mentioned the gauge dependent quantities are strongly correlated with the value of $F_{max}$.

The important question is how close we approach the global maximum. Theoretically SA algorithm permits to find the global maximum. In practice this is complicated task because of large autocorrelation length at small temperature $T$. We checked for a few configurations that increasing the number of sweeps in our SA algorithm by factor 5 does not lead to any essential change in $F_{max}$. This indicates that we are close to the equilibrium during SA cooling and thus our results are close to the global maximum when we make the extrapolation to $N_{cop} \to \infty$ limit.

As we show in the Appendix there are 2 possibilities for dependence of $< F_{max} >$ on $N_{cop}$. Our data show that at small $N_{cop}$ eq.(A.2) provides better fit while at large $N_{cop}$ eq.(A.1) is better. For this reason we use eq.(A.1) to extrapolate the data to $N_{cop} \to \infty$ limit. In Fig.2 we depict $\chi_{Z(2)}(5,5)$ and $< \rho >$ as functions of $< F_{max} >$. The linear dependence seen in this figure proves that the use of the same fitting function (eq. (A.1)) for $< F_{max} >$ and for $\chi_{Z(2)}(I,I)$ and $< \rho >$ is indeed well grounded.

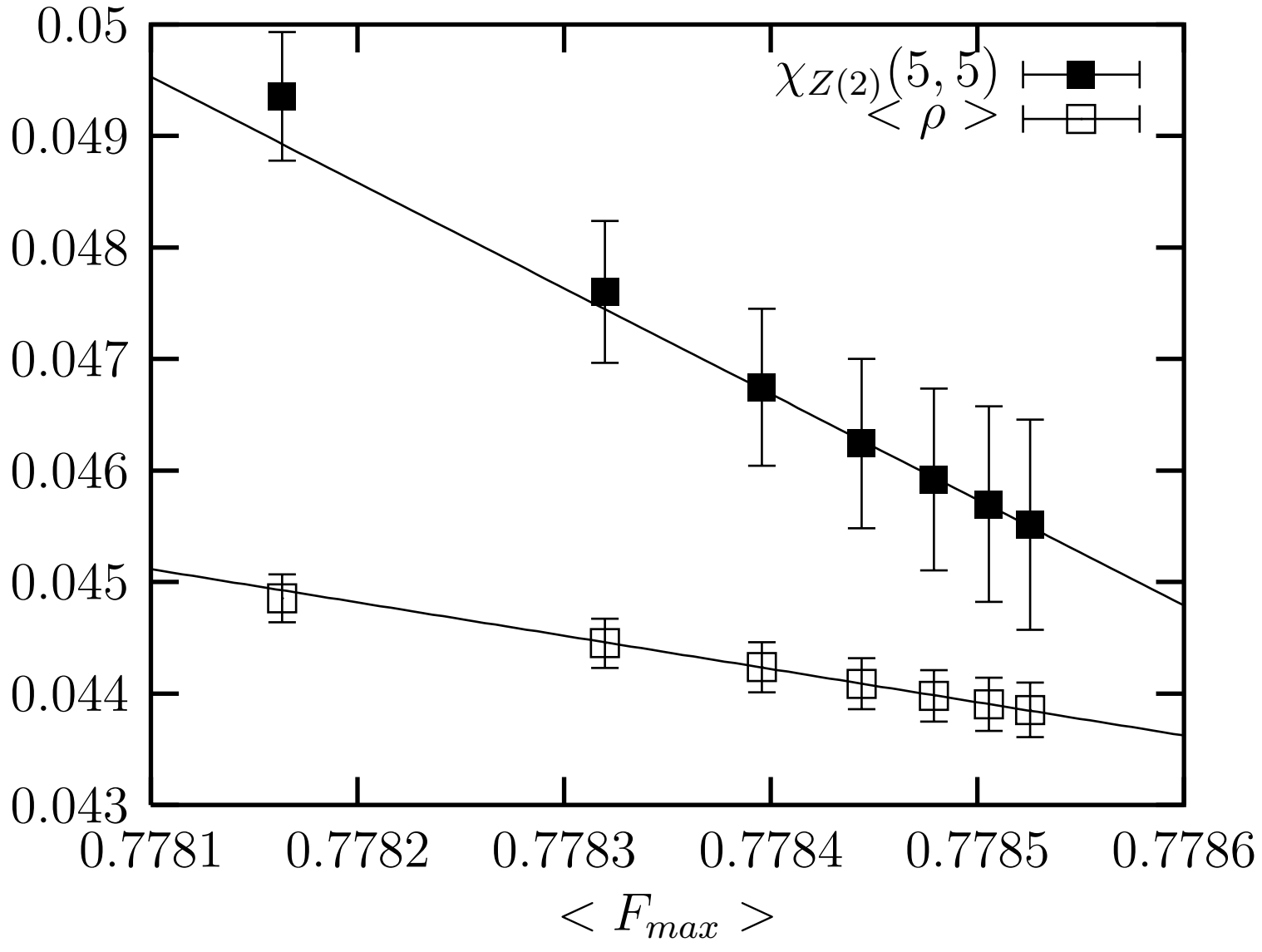

Figure 2: $\chi_{Z(2)}(5,5)$ and $< \rho >$ vs. $< F_{max} >$ for $L = 20, \beta = 2.4$.

## 3.3 P-vortex density

The physical density of P-vortices $< \rho/a^2 >$ is the important indicator of their relevance for the infrared physics. If, e.g., it goes to zero in the continuum limit, then P-vortices cannot play any important role. It is known that for the completely uncorrelated vortices $< \rho/a^2 >= \frac{1}{2}\sigma_{Z(2)}$ [13].

To compare the scaling of density with that of the unprojected string tension we show in Fig.3 the ratio[3] $<\rho>/\sigma_{SU(2)}a^2$. One can see that for two $\beta$ values this ratio is almost the same, i.e. scaling properties of $<\rho/a^2>$ and $\sigma_{SU(2)}$ are similar. The volume dependence is rather weak. Our value for $<\rho>$ is essentially lower than that obtained in [12] and this difference increases with $\beta$. It will be clear from our results in the next section that $<\rho/a^2>\approx\sigma_{Z(2)}$, i.e. P-vortices are far from being uncorrelated. On the other hand (at least on the lattices up to $24^4$) the density of P-vortices does not reproduce the full string tension, $\sigma_{SU(2)}$, as it was suggested in Ref. [4].

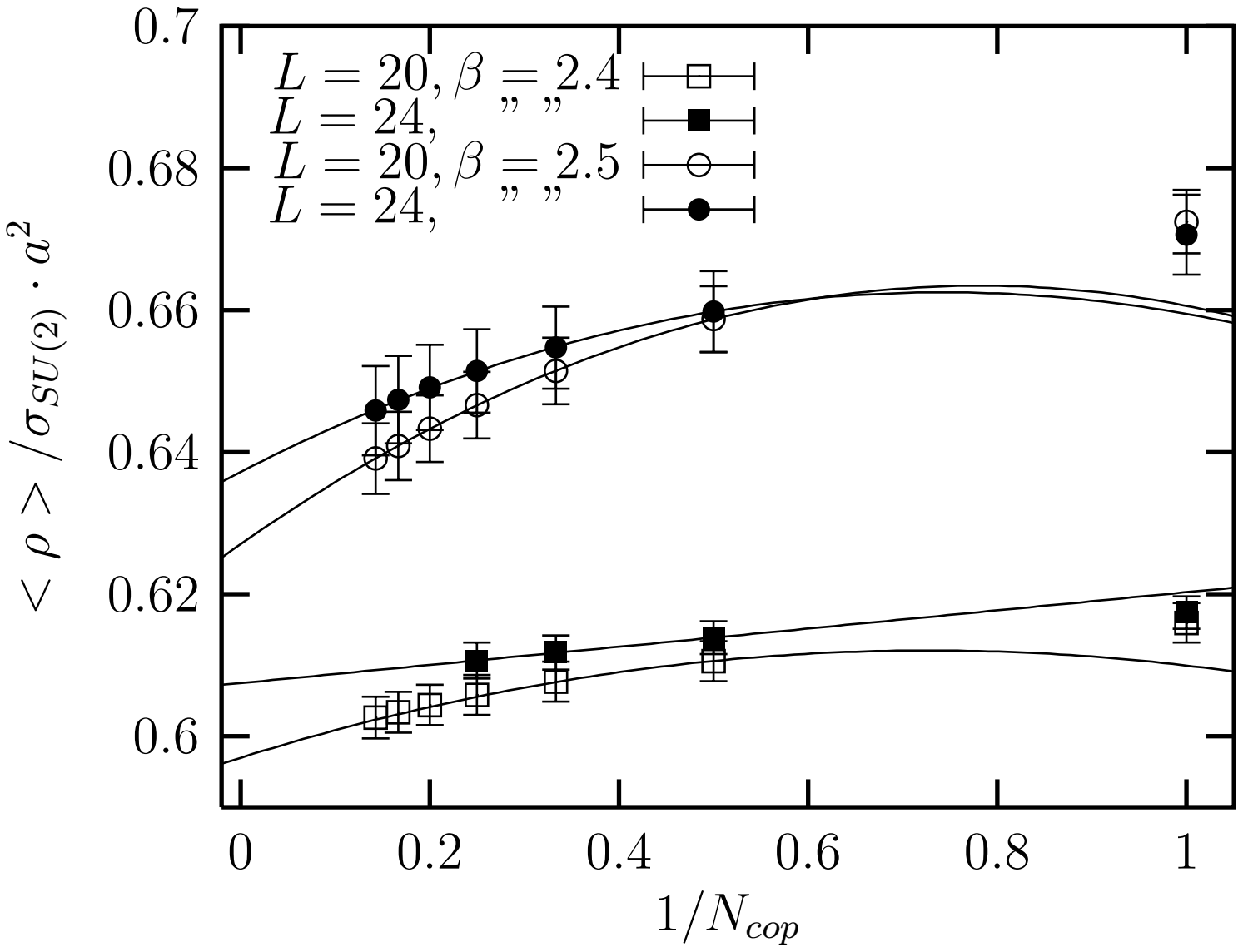

Figure 3: The ratio $<\rho>/\sigma_{SU(2)}a^2$ vs. $N_{cop}$. The solid lines are the fits obtained using asymptotic expression (A.1).

## 3.4 Projected string tension $\sigma_{Z(2)}$

### 3.4.1 Creutz ratio $\chi_{Z(2)}$

At first we discuss the final volume effects. In Fig.4 we show the dependence of $\chi_{Z(2)}(4,4)$ on $N_{cop}$ for $\beta=2.4$. From the figure one can see that finite volume effects are essential for $L=12$ only. For the other values of $I$ we obtained the similar behavior. At $\beta=2.5$ we found a sizable finite volume effects only on $L=16$ lattice.

[3] $\sigma_{SU(2)}a^2=0.0728(6)$ for $\beta=2.4$[23], $0.0350(4)$ for $\beta=2.5$ [24]

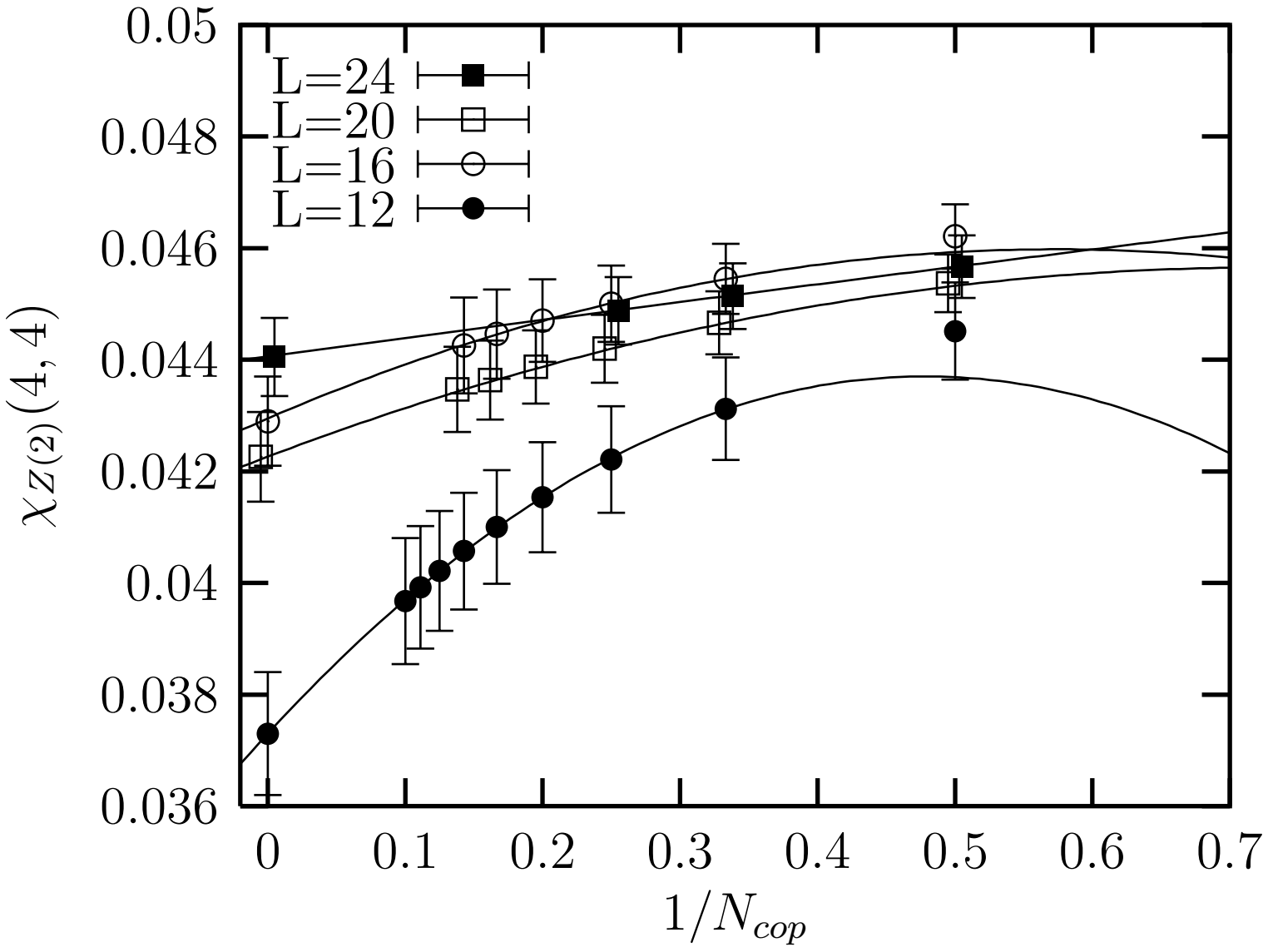

Figure 4: The dependence of the Creutz ratios $\chi_{Z(2)}(4,4)$ on the number of gauge copies $N_{cop}$ for $\beta = 2.4$. The solid lines are fits using eq. (A.1).

Usually one uses the Creutz ratio $\chi_{Z(2)}(I,I)$ (4) to estimate the projected string tension because the dependence on $I$ was found to be rather weak. In our study we found that it was not quite true. In Fig.5 we plot the ratio $\chi_{Z(2)}(I,I)/\sigma_{SU(2)}a^2$ as a function of the dimensional distance [4] $r = I \cdot a$. The data shown were obtained for maximal values of $N_{cop}$ given in Table 1. The values extrapolated to $N_{cop} \to \infty$ limit are only slightly lower. One can see that $\chi_{Z(2)}(I,I)$ is not constant but increases with $I$ at least in the range $0.3fm \lesssim I \cdot a \lesssim 0.7fm$. The similar behavior can be seen in Fig.3 of [12]. We found that this effect weakens, when $N_{cop}$ increases, but does not disappear completely. It is evident from this figure that even on our largest lattice ($L = 24$ at $\beta = 2.4$), which has extension in physical units of about $3fm$, $\chi_{Z(2)}(I,I)$ is much lower than the unprojected string tension. This invalidates the suggestion of [12] that $\chi_{Z(2)}(I,I)$ approaches $\sigma_{SU(2)}a^2$ on lattices larger than the size of $Z(2)$-vortex ($\sim 1fm$). Let us remind that the authors of Ref. [12] have used the RO algorithm to fix the gauge, while we are using the SA algorithm.

[4] The physical distance scale is set by the value $\sigma_{SU(2)} = (440\ Mev)^2$

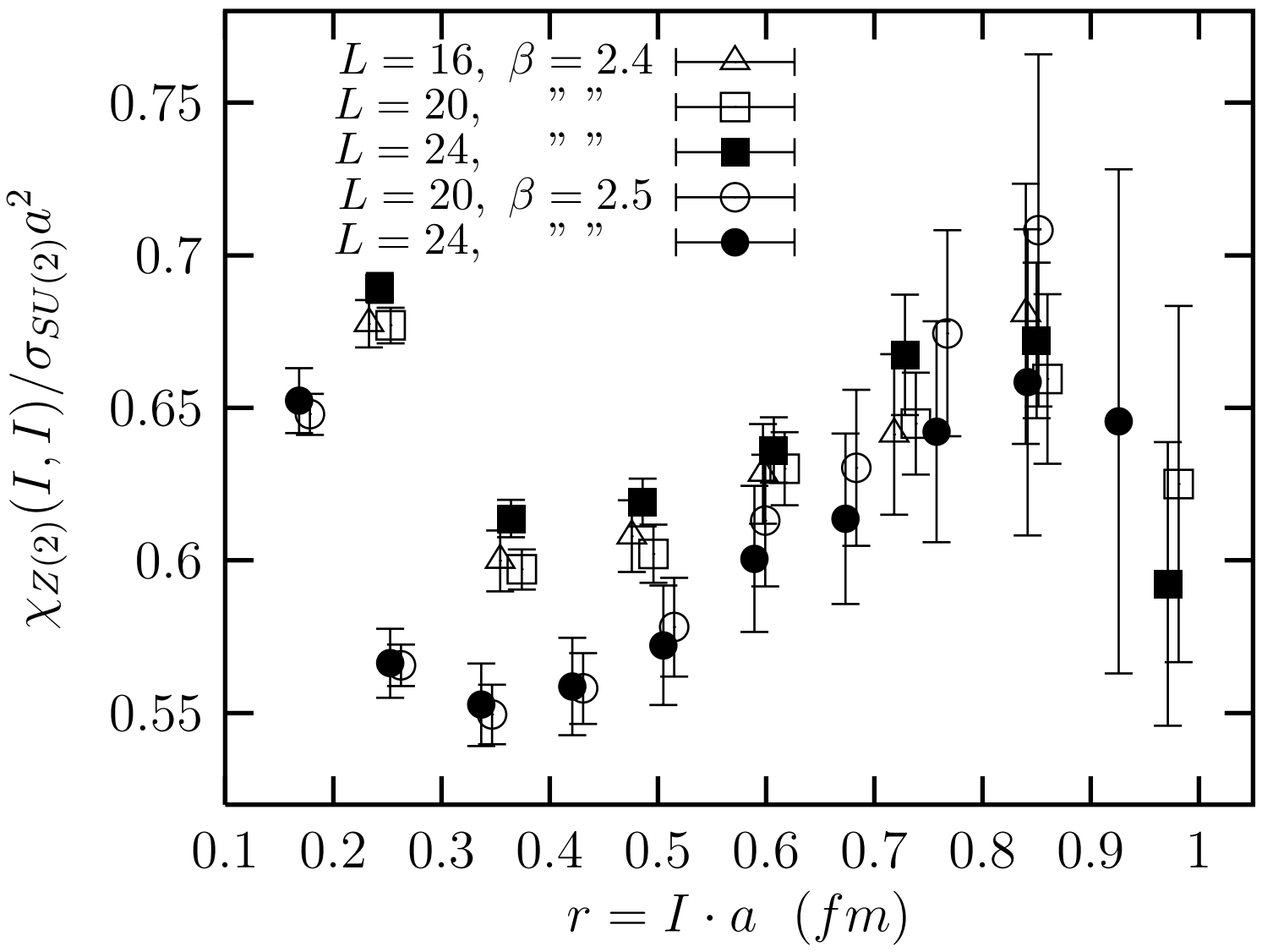

Figure 5: The dependence of $\chi_{Z(2)}(I,I)/\sigma_{SU(2)}a^2$ on the distance in physical units. $N_{cop}$ are the largest values used for the given lattice (see Table 1).

Our estimation for $\sigma_{Z(2)}/\sigma_{SU(2)}$ made from projected Creutz ratio measured at distances $0.7 - 0.9fm$ is 0.66(2). Comparing our data obtained at given $\beta$ on lattices of various size (see Fig.5), we can conclude that for lattices of physical extension larger than $1.7fm$ there are no sizable finite volume effects for the string tension extracted from the Wilson loops of the size $\lesssim 1fm$. Still we cannot exclude that the projected string tension, measured from projected Wilson loops with a very large extension, approaches the full string tension.

### 3.4.2 Static potential $V_{Z(2)}(R,T)$

In view of the dependence of $\chi_{Z(2)}(I,I)$ on $I$ it is interesting to study the projected static potential $V_{Z(2)}(R) = \lim_{T\to\infty} V_{Z(2)}(R,T)$, $V_{Z(2)}(R,T) = -\log \frac{W_{Z(2)}(R,T+1)}{W_{Z(2)}(R,T)}$. As it is seen from Fig.6 $V_{Z(2)}(R,T)$ rises as a function of $T$. If we assume that for $W_{Z(2)}(R,T)$ the usual expansion

$$W_{Z(2)}(R,T) = C_0(R)exp\left(-V_{Z(2)}(R)T\right) \ + \ C_1(R)exp\left(-V_1(R)T\right) \ + ... \qquad (6)$$

is valid, then such behavior implies that $C_1(R) < 0$.

To compute $V_{Z(2)}(R)$ we first extrapolated $V_{Z(2)}(R,T)$ as a function of $N_{cop}$ to $N_{cop} \to \infty$ limit and then evaluated $V_{Z(2)}(R)$ using large $T$ values. The results are depicted in Fig.7.

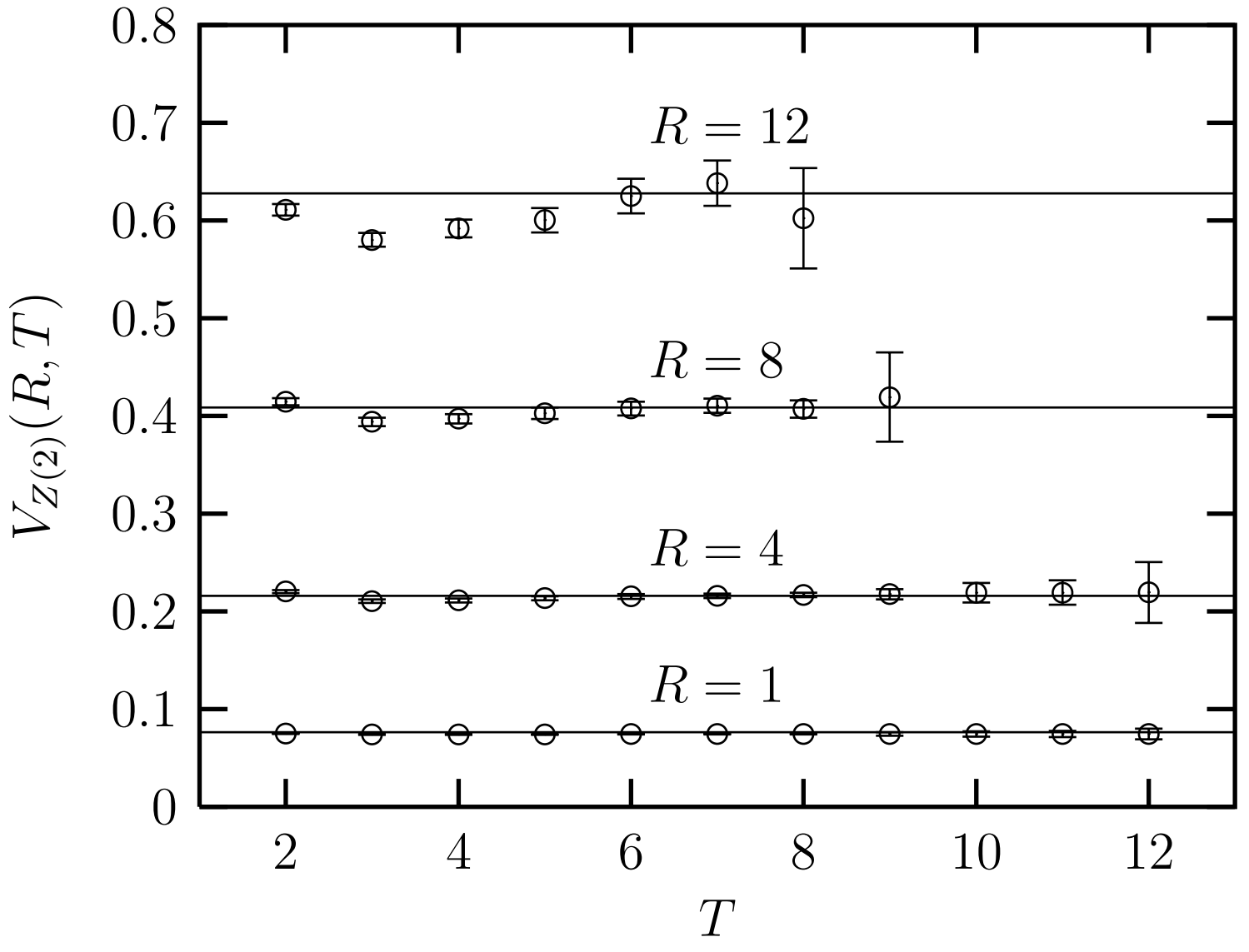

Figure 6: $V_{Z(2)}(R,T)$ as a function of $T$ for several values of $R$. $L = 24$, $\beta = 2.4$, $N_{cop} = 4$ .

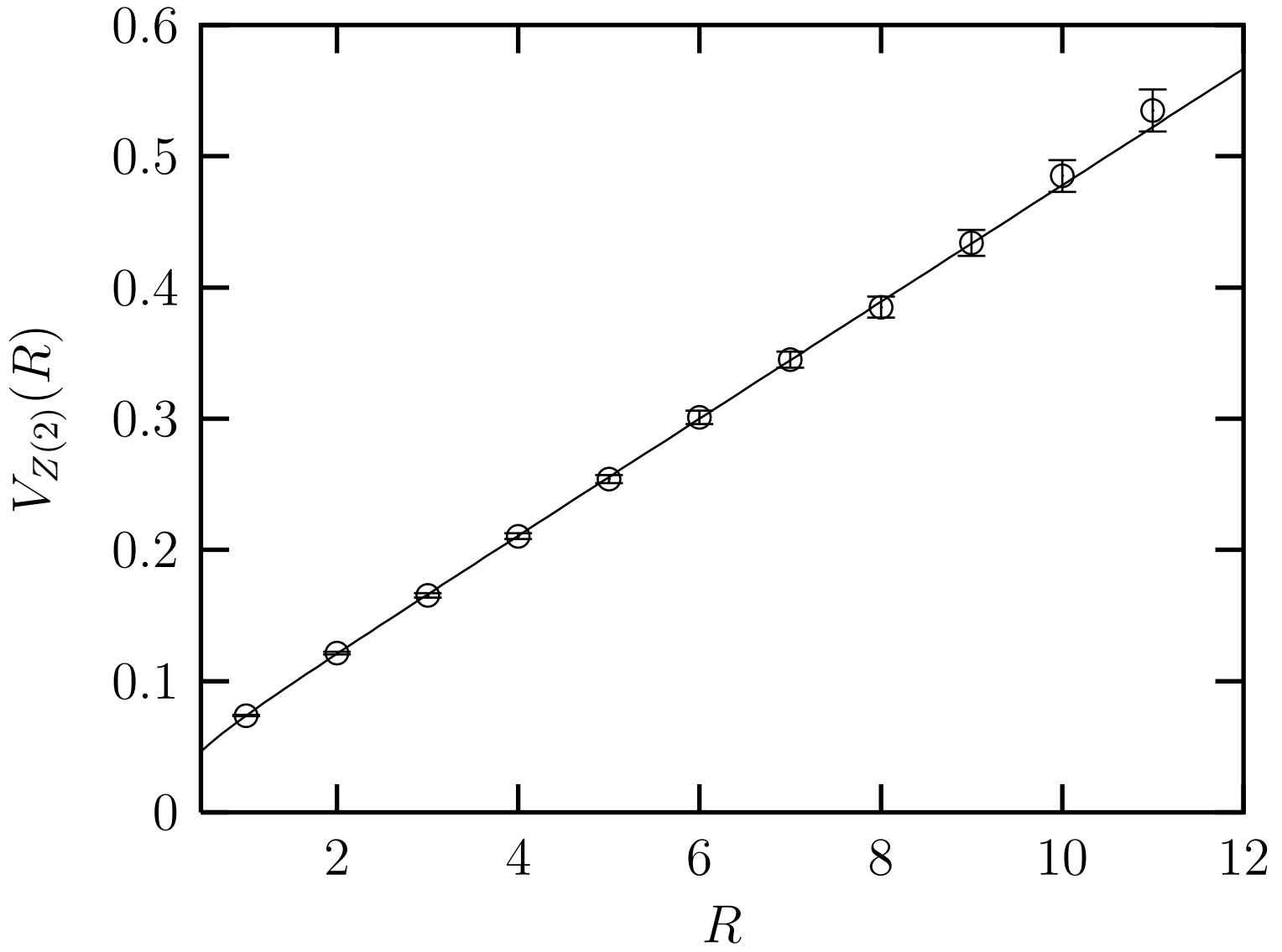

Figure 7: The static potential $V_{Z(2)}(R)$. $L = 24$, $\beta = 2.4$. The solid line is fit using eq.(5).

The value of $\sigma_{Z(2)}$ obtained from the projected static potential is in agreement with that obtained from the projected Creutz ratio (Section 3.4.1).

# 4 Conclusions

We made the most careful up to date gauge fixing of DMC gauge employing SA algorithm. We have shown that this algorithm permits to get the higher local maximum of the functional $F(U)$ than that obtained before with other algorithms. Though increment in $< F_{max} >$ looks not impressive it gives rise to substantial change in the values of gauge noninvariant observables. This, in turn, brought us to conclusions different from those made in [12]. We computed the projected string tension $\sigma_{Z(2)}$ and P-vortex density $< \rho >$ on volumes with physical extension from $1.4fm$ up to $3fm$. Our results imply that only on our smallest lattices ($L = 12$ for $\beta = 2.4$ and $L = 16$ for $\beta = 2.5$) there are moderate finite volume effects. Our data for lattice size larger than $1.7fm$ show the absence of the finite volume effects.

The ratio of P-vortex density to $\sigma_{SU(2)}$ weakly depends on $\beta$ (see Fig.3) and the early asymptotic scaling of this density (observed in [4, 12]) does not exist.

The projected string tension, $\sigma_{Z(2)}$, does not reproduce the full string tension, $\sigma_{SU(2)}$. The obtained value of the ratio $\sigma_{Z(2)}/\sigma_{SU(2)} \approx 0.66(2)$ is rather low. Comparing results for two lattice spacings we do not see any convergence of this ratio to 1 in the continuum limit (further studies at larger values of $\beta$ on larger lattices are important). Let us remind that abelian monopoles are responsible for $92\% - 94\%$ of the full $SU(2)$ string tension [17, 25].

Since the P-vortices in DMC gauge do not reproduce the full $SU(2)$ string tension they cannot be responsible for confinement. On the other hand the fact that $\sigma_{Z(2)}$ is of the same order as $\sigma_{SU(2)}$ probably means that the DMC gauge is close to some "ideal gauge" which detect perfectly [3] the thick gauge independent center vortices [26]. Recent results [27] show that an example of such gauge is the Laplacian Center Gauge which is free of Gribov copies problem and $\sigma_{Z(2)} \approx \sigma_{SU(2)}$ in this gauge.

# Acknowledgements

We are grateful to V. Anikeeva and A. Veselov for the help at some stages of this work. We thank T. Kovacs for the helpful remark. This work was partially supported by grants RFBR 99-01230a, INTAS 96-370, Monbushu grant and CRDF award RP1-2103.

# Appendix A

Let the probability distribution of the random variable $x$ is $p(x)$ and let $p(x)$ is nonzero on the interval $[x_0-\delta_1, x_0+\delta_2]$. Let us introduce

$$P(x) = \int_{x_0-\delta_1}^{x} dy\ p(y)$$

with properties

$$P'(x) = p(x), \quad P(x_0-\delta_1) = 0, \quad P(x_0+\delta_2) = 1.$$

Then probability distribution $P^{(N)}(x)$ to get value x as maximal out of N attempts is equal to

$$P^{(1)}(x) = p(x)$$

$$P^{(N)}(x) = C_N P(x)^{N-1} p(x)$$

where normalization constant

$$C_N^{-1} = \int_{x_0-\delta_1}^{x_0+\delta_2} dx\ P(x)^{N-1} p(x) = \int_{P(x_0-\delta_1)}^{P(x_0+\delta_2)} dP\ P^{N-1} = \frac{1}{N} P^N|_0^1 = \frac{1}{N}$$

We have to evaluate

$$\bar{x}^{(N)} = \int_{x_0-\delta_1}^{x_0+\delta_2} dx\ xP^{(N)}(x) = N \int_{x_0-\delta_1}^{x_0+\delta_2} dx\ xp(x)P^{N-1}(x)$$

$$= x_0 + N \int_{-\delta_1}^{\delta_2} dy\ yp(y)P^{N-1}(y) \quad (x = y + x_0)$$

$$= x_0 + \int_{-\delta_1}^{\delta_2} dy\ y \frac{dP^N(y)}{dy} = x_0 + yP^N(y)|_{-\delta_1}^{\delta_2} - \int_{-\delta_1}^{\delta_2} dy\ P^N(y)$$

$$= x_0 + \delta_2 - \int_{-\delta_1}^{\delta_2} dy\ P^N(y) = \bar{x} - \int_{-\delta_1}^{\delta_2} dy\ P^N(y)$$

where $\bar{x} = x_0 + \delta_2$ is evidently the value of $\bar{x}^{(N)}$ in the limit $N \to \infty$. The integral in the last equation can be evaluated for large $N$ in the following way:

$$I = -\int_{-\delta_1}^{\delta_2} dy\ P^N(y) = -\int_{-\delta_1}^{\delta_2} dy\ e^{Nlog(P(y))}$$

We expand the function $f(y) = log(P(y))$ at its maximum $y = \delta_2$:

$$f(y) = \frac{P'(\delta_2)}{P(\delta_2)}\epsilon + \frac{1}{2}[\frac{P''(\delta_2)}{P(\delta_2)} - (\frac{P'(\delta_2)}{P(\delta_2)})^2]\epsilon^2 + ... = p(\delta_2)\epsilon + \frac{1}{2}[p'(\delta_2) - p^2(\delta_2)]\epsilon^2 + ..., \quad \epsilon = y - \delta_2$$

Then

$$I = -\int_{-\delta_1-\delta_2}^{0} d\epsilon\ e^{N\left(p(\delta_2)\epsilon + \frac{1}{2}[p'(\delta_2)-p^2(\delta_2)]\epsilon^2 + ...\right)}$$

Let us consider 2 cases:

1. $p(\delta_2) \neq 0$.

$$I = -\frac{1}{N}\int_{-N(\delta_1+\delta_2)}^{0} d\epsilon\ e^{p(\delta_2)\epsilon + \frac{1}{2N}[p'(\delta_2)-p^2(\delta_2)]\epsilon^2 + ...} = -\frac{1}{Np(\delta_2)}\left(1 + O\left(\frac{1}{N}\right)\right)$$

Thus in this case the fitting function for numerical data should be:

$$f_1(N) = \bar{x} + \frac{C_1}{N} + \frac{C_2}{N^2} + ... \tag{A.1}$$

2. $p(\delta_2) = 0$.

$$I = -\frac{1}{\sqrt{N}}\int_{-\sqrt{N}(\delta_1+\delta_2)}^{0} d\epsilon\ e^{\frac{1}{2}[p'(\delta_2)-p^2(\delta_2)]\epsilon^2 + ...}$$

Then fitting function is

$$f_2(N) = \bar{x} + \frac{C_1}{\sqrt{N}} + \frac{C_2}{N} + ... \tag{A.2}$$

If $p(\delta_2)$ is a very small number then there might exist interval of $N$ where eq. (A.2) provides better fit, while in the limit of large $N$ eq.(A.1) is valid. The data for our observables ($< F_{max} >$, $< \rho >$, $\chi_{Z(2)}(I, I)$) indicates just this situation. For this reason we made fits with eq.(A.1) discarding small values of $N_{cop}$.

# References

[1] G. 't Hooft, *Nucl. Phys.* **B138** ,1 (1978);
J.M. Cornwall, *Nucl. Phys.* **B157** ,392 (1979);
G. Mack, in Recent Developments in Gauge Theories, ed. by G. 't Hooft et al. (Plenum, New York, 1980);
H.B. Nielsen and P. Olesen, *Nucl. Phys.* **B160** (1979) 380;
J. Ambjorn and P. Olesen, *Nucl. Phys.* **B170** (1980) 60; 265.

[2] T.G. Kovacs and E. T. Tomboulis, *J.Math.Phys.* **40** (1999) 4677.

[3] L. Del Debbio, M. Faber, J. Greensite and S. Olejnik, *Phys.Rev.* **D55** (1997) 2298.

[4] L. Del Debbio, M. Faber, J. Giedt, J. Greensite and S. Olejnik, *Phys.Rev.* **D58** (1998) 094501.

[5] Ph. de Forcrand and M. D'Elia, *Phys.Rev.Lett.* **82** (1999) 4582;
C. Alexandrou, M. D'Elia and Ph. de Forcrand, *Nucl.Phys.* **B** [Proc.Suppl.] **83**(2000) 437.

[6] B.L. Bakker, A.I. Veselov and M.A. Zubkov, hep-lat/007022.

[7] V.N. Gribov, *Nucl. Phys.* **B139** (1978) 1.

[8] T.G. Kovacs and E.T. Tomboulis, *Phys.Lett.* **B463** (1999) 104.

[9] V.G. Bornyakov, D.A. Komarov, M.I. Polikarpov and A.I. Veselov, *JETP Lett.* **71** (2000) 231.

[10] V.G. Bornyakov, D.A. Komarov, M.I. Polikarpov and A.I. Veselov, To be published in Proceedings of the International Symposium on Quantum Chromodynamics and Color Confinement (Confinement 2000), Osaka, Japan, 7-10 Mar 2000.

[11] J. Greensite, hep-lat/0005001.

[12] R. Bertle, M. Faber, J. Greensite, S. Olejnik, hep-lat/0007043.

[13] K. Langfeld, O. Tennert, M. Engelhardt and H. Reinhardt, *Phys.Lett.* **B452** (1999) 301.

[14] J.E. Mandula and M.C. Ogilvie, *Phys. Lett.* **185B** (1987) 127.

[15] H. Nakajima, S. Furui and A. Yamaguchi, hep-lat/0007001.

[16] A. Nakamura and M. Plewnia, *Phys. Lett.* **255B** (1991) 274;
V. Bornyakov, V.K. Mitrjushkin, M. Müller-Preussker and F. Pahl, *Phys. Lett.* **B317B** (1993) 596.

[17] G.S. Bali, V. Bornyakov, M. Müller-Preussker and K. Schilling, *Phys.Rev.* **D54** (1996) 2863.

[18] F. Shoji, T. Suzuki, H. Kodama and A. Nakamura, *Phys. Lett.* **476B** (2000) 199.

[19] G. Damm, W. Kerler, *Phys. Lett.* **397B** (1997) 216.

[20] S. Fachin and C. Parrinello, *Nucl. Phys.* **B** [Proc. Suppl.] **26**(1992) 429.

[21] J.C. Vink, U.-J. Wiese, *Phys.Lett.* **B289** (1992) 122;
A.J. van der Sijs, *Prog.Theor.Phys.Suppl.* **131** (1998) 149.

[22] S. Kirkpatrick, C.D. Gelatt Jr., M.P. Vecchi, *Science* **220** (1983) 671; V. Cerny, *J. Opt. The. Appl.* **45** (1985) 41.

[23] S. Perantonis, A. Huntley and C. Michael, *Nucl. Phys.* **B326** (1989) 544.

[24] G.S. Bali, K. Schilling and C. Schlichter, *Phys.Rev.* **D51** (1995) 5165.

[25] F.V. Gubarev, E.M. Ilgenfritz, M.I. Polikarpov and T. Suzuki, *Phys. Lett.* **B468** (1999) 134.

[26] J. Greensite, private communication.

[27] Ph. de Forcrand and M. Pepe, hep-lat/0008016.